\documentclass[pre,aps,reprint,amsmath]{revtex4-1}

\usepackage[T1]{fontenc} 
\usepackage{graphicx}
\usepackage[utf8]{inputenc}
\usepackage{lmodern}
\usepackage{bm}
\usepackage{bbold}
\usepackage[hidelinks]{hyperref}				
\usepackage[dvipsnames]{xcolor}
\usepackage{ulem}
\newcommand{\ms}{\color{black}}

\newcommand{\rv}{{\bf r}}
\newcommand{\Fv}{{\bf F}}
\newcommand{\fv}{{\bf f}}
\newcommand{\Gv}{{\bf G}}
\newcommand{\Jv}{{\bf J}}
\newcommand{\vel}{{\bf v}}
\newcommand{\type}{e}

\begin{document}
\title{\ms Universality in Driven and Equilibrium Hard Sphere Liquid Dynamics}

\author{Lucas L. Treffenst\"adt}
\affiliation{Theoretische Physik II, Physikalisches Institut,
  Universit{\"a}t Bayreuth, D-95447 Bayreuth, Germany}

\author{Matthias Schmidt}
\affiliation{Theoretische Physik II, Physikalisches Institut,
  Universit{\"a}t Bayreuth, D-95447 Bayreuth, Germany}
\email{Matthias.Schmidt@uni-bayreuth.de}

\date{10 October 2020, to appear in Phys. Rev. Lett.}

\begin{abstract}
We demonstrate that the time evolution of the van Hove dynamical pair
correlation function is governed by adiabatic forces that arise from
the free energy and by superadiabatic forces that are induced by the
flow of the van Hove function. The superadiabatic forces consist of
drag, viscous, and structural contributions, as occur in active
Brownian particles, in liquids under shear and in lane forming
mixtures. For hard sphere liquids we present a power functional theory
that predicts these universal force fields in quantitative agreement
with our Brownian dynamics simulation results.
\end{abstract}

\maketitle

The van Hove function is arguably one of the most fundamental
correlators that characterize the dynamical pair structure of a liquid
on the microscopic scale \cite{vanhove1954,hansen2013}.
{\ms It measures the probability of finding two particles at a
  distance $r$, where the particles are randomly chosen, but with a
  time lapse of duration~$t$ between the two position measurements.
  Both the motion of the same particle, as well as spatio-temporal
  correlations of two distinct particles are captured.}
Significant physical insights into the dynamics of {\ms both simple
  and} complex systems {\ms could be} gained from studying their van
Hove function. Examples thereof including cage formation in nematic
and smectic liquid crystals \cite{bier2008prl}, {\ms de Gennes
  narrowing of liquid iron \cite{wu2018}, self-motion of water
  \cite{shinohara2020},} the dynamics of colloidal hard spheres
\cite{stopper2015pre,stopper2015jcpCommunication, stopper2018exp} and
of colloid-polymer mixtures \cite{stopper2016jpcm}.  Experimentally,
highly accurate results for the van Hove function are accessible based
on microscopy of colloidal systems~\cite{stopper2018exp}, {\ms as well
  as by scattering methods, which yield the Fourier
  transform~\cite{hansen2013,shinohara2020}.}

Much of our knowledge and understanding of the properties of the van
Hove function is based on computer simulation work. Formulating a
theoretical description for the complex spatial and temporal two-body
dynamics remains a formidable challenge. Much insightful work has been
carried out by Medina-Noyola and his coworkers on the basis of
generalized Langevin equations \cite{yeomansreyna2000,chavezrojo2006,
  lopezflores2012,lopezflores2013,lazarolazaro2017,
  yeomans-reyna2003,yeomans-reyna2007, juarez-maldonado2007,
  medina-noyola2009,ramirez-gonzales2010}, but also mode-coupling
theory was used at high densities~\cite{weysser2010}.  Furthermore,
the closely related problem of identifying and studying memory kernels
has received much recent attention in the context of of molecular
dynamics~\cite{lesnicki2016,lesnicki2017,jung2016,jung2017,jung2018}.

The dynamical test particle limit
\cite{archer2007dtpl,hopkins2010dtpl,brader2014dtpl} constitutes a
formally exact reformulation of the time evolution of the van Hove
function in a one-body picture. Instead of working explicitly with
two-body correlations, an equivalent dynamical situation is
constructed, where one-body profiles evolve in time, which offers
significant conceptual simplification. Fixing a particle at the
initial time at the origin establishes the equivalence with the
original problem. The concept is formally exact, but it requires a
prescription for the one-body dynamics to be useful in practice.

When choosing the dynamical density functional theory (DDFT)
\cite{evans1979,marconi1999,archer2004} to perform the one-body
dynamics of the van Hove function
\cite{archer2007dtpl,hopkins2010dtpl,
  stopper2015pre,stopper2015jcpCommunication, stopper2018exp} one
finds too rapid temporal decay of the interparticle correlations
\cite{reinhardt2012}, as compared to benchmark data from Brownian
dynamics (BD) computer simulations. This trend persists even when
choosing Rosenfeld's fundamental measure theory
\cite{rosenfeld1989,tarazona2008review,roth2010review,lutsko2010review}
as an excellent approximation for the (hard sphere) free energy
functional. Accounting for the observed reduction of particle mobility
at increased density requires empirical adjustments to the DDFT
framework \cite{hopkins2010dtpl,
  stopper2015pre,stopper2015jcpCommunication, stopper2018exp}.

Power functional theory (PFT) \cite{schmidt2013pft} provides formally
exact test particle dynamics \cite{brader2014dtpl}, albeit very little
explicit knowledge of the crucial superadiabatic force contributions
\cite{schmidt2013pft,fortini2014prl}, i.e.\ those beyond DDFT, had
originally been available \cite{brader2014dtpl}. In BD simulation work
it was shown that the superadiabatic forces that govern the van Hove
function are both significant in magnitude and nontrivial in their
spatial and temporal structure \cite{schindler2016dtpl}.  In a variety
of {\it nonequilibrium} systems, different superadiabatic force types
were identified as providing the key mechanisms for prominent physical
effects, such as the emergence of viscous and structural forces in BD
flow
\cite{delasheras2018velocityGradient,stuhlmueller2018prl,delasheras2020prl},
motility-induced phase separation in active Brownian particles
\cite{krinninger2016prl,hermann2019prl}, spontaneous lane formation in
counter-driven mixtures \cite{geigenfeind2020superdemixing}, and
memory-induced motion reversal \cite{treffenstaedt2020shear}.

Here we show that the identical types of superadiabatic forces that
rule the behaviour of these driven systems, determine both
qualitatively and quantitatively the van Hove function, and hence the
intrinsic equilibrium dynamics. That the same form of superadiabatic
forces apply across such a wide range of different physical situations
indicates that the microscopic liquid dynamics are governed by
universal mechanisms. Besides the conceptual importance of this
finding, it allows concrete cross fertilization between results
obtained for apparently very different systems.

\begin{figure*}
  \includegraphics[width=0.95\textwidth,angle=0]{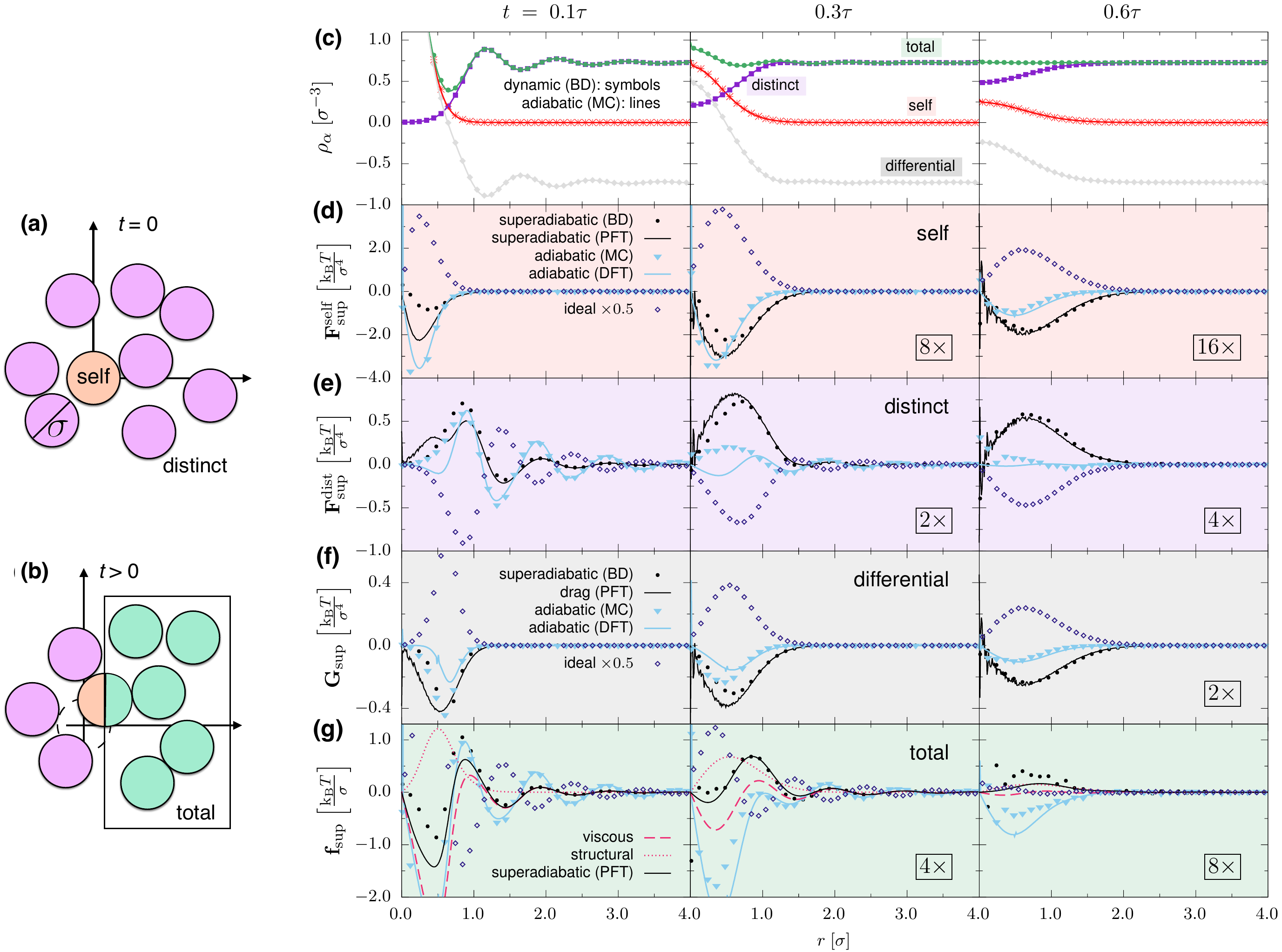}
  \caption{{\bf (a)} Illustration of the van Hove dynamical two-body
    correlation function in a bulk liquid of hard spheres of
    diameter~$\sigma$ at time $t=0$ and {\bf (b)} and $t>0$.
    {\bf (c)}-{\bf (g)} Results for the dynamical decay of the
    two-body structure of a bulk liquid of hard spheres at packing
    fraction 0.35 at times $t=0.1\tau$ (left column), $0.3\tau$
    (middle column), and $0.6\tau$ (right column) and as a function of
    the scaled distance $r/\sigma$.
    {\bf(c)} Total ($\rho$), self ($\rho_{\rm self}$), distinct
    ($\rho_{\rm dist}$), and differential ($\rho_\Delta$) parts of the
    van Hove function, as indicated. The results from BD simulation
    (symbols) of the time evolution and from MC simulation (lines) of
    the corresponding adiabatic state coincide on the scale of the
    plot.
    {\bf (d)} Self part of the superadiabatic force density $\Fv_{\rm
      sup}^{\rm self}$, as obtained from BD (symbols) and from PFT
    (solid line); also shown is the adiabatic self force density
    $\Fv_{\rm ad}^{\rm self}$ from MC (blue symbols) and DFT (blue
    line).  The ideal self force density, $-k_BT\nabla\rho_{\rm
      self}$, is shown as a reference.
    {\bf (e)} Distinct part of the superadiabatic force density, $\Fv_{\rm
      sup}^{\rm dist}$, as obtained from BD and from PFT.
    {\bf (f)} Differential superadiabatic (drag) force density
    $\Gv_{\rm sup}$, as obtained from BD (symbols) and from PFT (black
    line), and differential adiabatic force density $\Gv_{\rm ad}$, as
    arising from the adiabatic self correction.
    {\bf (g)} Species-independent superadiabatic force field $\fv_{\rm
      sup}$ as obtained from BD and from PFT, along with the
    theoretical viscous ($\fv_{\rm visc}=-\rho^{-1}\delta P_t^{\rm
      visc}/\delta \vel$) and structual contributions ($\fv_{\rm
      struc}=-\rho^{-1}\delta P_t^{\rm struc}/\delta\vel$), where
    $\fv_{\rm sup}=\fv_{\rm visc}+\fv_{\rm struc}$.
    For the sake of clarity, the results in (d)-(g) at $t=0.3\tau$ and
    $0.6\tau$ are multiplied by a factor of $2\times, 4\times,
    8\times$ or $16\times$, as indicated. }
  \label{figResults}
\end{figure*}

Within the dynamical test particle limit, the van Hove function is
expressed as a time-dependent one-body density profile $\rho(\rv,t)$;
here $\rv$ indicates position and $t$ time. Often one splits into self
and distinct parts, $\rho(\rv,t)=\rho_{\rm self}(\rv,t)+\rho_{\rm
  dist}(\rv,t)$.  At the initial time the test (``self'') particle is
taken to be at the origin and the distinct particles are distributed
according to the (static) pair correlation function $g(r)$ of the bulk
liquid (as prescribed by Percus' static test particle limit
\cite{percus1962}). Hence the initial conditions, at $t=0$, are
$
  \rho_{\rm self}(\rv,0) = \delta(\rv)
$ and $
  \rho_{\rm dist}(\rv,0) = \rho_b g(r), 
$ 
where $\delta(\cdot)$ indicates the Dirac delta function, $\rho_b$ is
the bulk fluid number density, and
$r=|\rv|$. Figs.~\ref{figResults}(a) and (b) depict an
illustration. The dynamics of the van Hove function are associated
with time-dependent one-body self and distinct currents, $\Jv_{\rm
  self}(\rv,t)$ and $\Jv_{\rm dist}(\rv,t)$, respectively. The total
van Hove current is the sum $\Jv=\Jv_{\rm self}+\Jv_{\rm dist}$. A
continuity equation holds for each species: $\partial
\rho_\alpha/\partial t = -\nabla\cdot \Jv_\alpha$, where $\alpha=\rm
self$, $\rm dist$ labels the two different species and $\nabla$
indicates the derivative with respect to $\rv$. The partial one-body
currents $\Jv_\alpha(\rv,t)$ arise both from free diffusion and from
internal interactions. Hence the one-body force density balance
relation is
\begin{equation}
  \gamma \Jv_\alpha = - k_BT \nabla \rho_\alpha + \Fv_{\rm int}^\alpha,
  \label{EQcurrentForceBalancePartial}
\end{equation}
where $\gamma$ is the friction constant against the static background,
$k_B$ is the Boltzmann constant, $T$ indicates absolute temperature,
and $\Fv_{\rm int}^\alpha(\rv,t)$ is the internal force density
distribution that acts on species $\alpha$. No external forces act in
the bulk system, and the time-dependent (nonequilibrium) situation is
solely introduced by the initial conditions $\rho_\alpha(\rv,0)$.

The internal force density consists of adiabatic ($\Fv_{\rm
  ad}^\alpha$) and superadiabatic ($\Fv_{\rm sup}^\alpha$)
contributions \cite{schmidt2013pft,fortini2014prl}, according to the
sum
$  
  \Fv_{\rm int}^\alpha = \Fv_{\rm ad}^\alpha + \Fv_{\rm sup}^\alpha.
$ 
Here $\Fv_{\rm ad}^\alpha(\rv,t)$ is the force density distribution in
an equilibrium (``adiabatic'') system that is defined to possess
one-body density profiles $\rho_{\rm ad}^\alpha(\rv)$ that are
identical to those in the dynamical system at time $t$: $\rho_{\rm
  ad}^\alpha(\rv)\equiv\rho_\alpha(\rv,t)$. This adiabatic
construction is performed at each point in time, and hence $\rho_{\rm
  ad}^\alpha(\rv)$ depends parametrically on $t$. The interparticle
interaction potential in the adiabatic system is identical to that in
the original dynamical system. The density distributions $\rho_{\rm
  ad}^\alpha(\rv)$ in the adiabatic system are stabilized by
species-dependent external potentials $V_{\rm ad}^\alpha(\rv)$, which
are guaranteed to exist in the adiabatic system due to the
Mermin-Evans map of classical density functional theory (DFT)
\cite{hansen2013,evans1979}.

The adiabatic force density can either be obtained by direct sampling
in the adiabatic system \cite{delasheras2019customFlow}, or, as we do
here, from the force balance in the adiabatic system:
\begin{equation}
  \Fv_{\rm ad}^\alpha 
  = k_BT\nabla\rho_\alpha+\rho_\alpha\nabla V_{\rm ad}^\alpha,
  \label{EQFadFromEquilibriumForceBalance}
\end{equation}
where all quantities on the right hand side are known.  In practice we
use a variant of the custom flow iterative method
\cite{delasheras2019customFlow}, where we sample the density profile
at each iteration step using Monte Carlo, and adjust the external
potentials $V_{\rm ad}^\alpha(\rv)$ accordingly \cite{fortini2014prl}
until the sampled density profiles in the adiabatic system match the
dynamical (``target'') density profiles $\rho_\alpha(\rv,t)$.

Within classical density functional theory the adiabatic internal
force density acting on species $\alpha$ is given by
$ \Fv_{\rm ad}^\alpha(\rv,t) = 
  -\rho_\alpha(\rv,t)
  \nabla \delta F_{\rm exc}/\delta\rho_\alpha(\rv,t),
$
where $F_{\rm exc}$ is the excess (over ideal gas) intrinsic Helmholtz
free energy functional. For the case of hard spheres, Rosenfeld's
fundamental measure theory
\cite{rosenfeld1989,roth2010review,tarazona2008review,lutsko2010review}
constitutes an excellent approximation for $F_{\rm exc}$. We
furthermore use the ``quenched'' approach by Stopper et
al.~\cite{stopper2015jcpCommunication}, where a self-correction is
applied in order to account for the fact that the self density profile
represents a single particle sharply (rather than a grand ensemble
average). This approach avoids having to use canonical decomposition
\cite{delasheras2014canonical,delasheras2016particleConservation} in
order to generate results that are specific to fixed particle number.

Within PFT the superadiabatic force density is obtained from a
functional derivative of the superadiabatic excess free power
functional $P_t^{\rm exc}$ according to
\begin{equation}
  \Fv_{\rm sup}^\alpha(\rv,t) =
  -\frac{\delta P_t^{\rm exc}}{\delta \vel_\alpha(\rv,t)},
  \label{EQFsupFromFunctional}
\end{equation}
where the derivative is taken at fixed density profiles, and the
species-resolved one-body velocity profile is
$\vel_\alpha(\rv,t)=\Jv_\alpha(\rv,t)/\rho_\alpha(\rv,t)$.  As an
approximation we use a functional that consists of drag
\cite{krinninger2016prl,hermann2019prl,geigenfeind2020superdemixing},
viscous \cite{delasheras2018velocityGradient,stuhlmueller2018prl,
  geigenfeind2020superdemixing,treffenstaedt2020shear,
  delasheras2020prl}, and structural
\cite{stuhlmueller2018prl,hermann2019prl,delasheras2020prl}
contributions, $P_t^{\rm exc}=P_t^{\rm drag}+P_t^{\rm visc}+P_t^{\rm
  struc}$, according to
\begin{eqnarray}
  P_t^{\rm exc} &=&
  \frac{C_{\rm drag}}{2}
  \int d\rv \rho_{\rm self}\rho_{\rm dist}
  (\vel_{\rm self}-\vel_{\rm dist})^2
  \label{EQPtexc}\\&&\quad
  +\int d\rv d\rv' \int_0^t dt' n_3 n_3'
  (\nabla\cdot\vel) (\nabla'\cdot\vel') K_{\rm visc}
  \notag \\&&\quad
    -\int d\rv d\rv'\int_0^tdt'(n_3' \vel')^2 (\nabla\cdot\Jv)
    K_{\rm struc},
\notag
\end{eqnarray}
where $C_{\rm drag}$ is a constant and the kernels $K_\type(\Delta
\rv, \Delta t)$, where $\type={\rm visc}, {\rm struc}$, depend on the
relative spatial and temporal distances $\Delta \rv=\rv-\rv'$ and
$\Delta t=t-t'$; the local packing fraction $n_3(\rv,t)$ is obtained
by convolution with $\rho(\rv,t)$
\cite{rosenfeld1989,tarazona2008review,roth2010review,lutsko2010review};
$\nabla'$ indicates the derivative with respect to $\rv'$, and we use
the shorthand $n_3'\equiv n_3(\rv',t')$ and
$\vel'\equiv\vel(\rv',t')$.  Here the total microscopic velocity
profile is $\vel=\Jv/\rho$.  We use the diffusing memory form
\cite{treffenstaedt2020shear} for $K_\type(\Delta\rv,\Delta t)$, which
consists of a product of a constant $C_\type$ that controls the
overall strength, an exponential decay with decay time constant
$\tau_\type$ and a diffusing Gaussian with diffusion constant
$D_\type$. Explicitly the form is
\begin{equation}
  K_\type(\Delta \rv,\Delta t) = 
  \frac{C_\type \exp(-\Delta \rv^2/(4 D_\type \Delta t)
    -\Delta t/\tau_\type)}
  {(4\pi D_\type \Delta t)^{3/2}\tau_\type}.
\end{equation}
The derivative \eqref{EQFsupFromFunctional} when applied to
\eqref{EQPtexc}, yields an explicit expression for $\Fv_{\rm
  sup}^\alpha$, which we evaluate below, using BD data for
$\rho_\alpha$ and $\vel_\alpha$ as input.  We choose the following set
of parameters: The drag strength is $C_{\rm drag} =
2.2\gamma\sigma^3$. The values for viscous memory kernel are identical
to those used in Ref.~\cite{treffenstaedt2020shear}: $C_{\rm visc} =
5.8 k_B T / ( \sigma^3 \tau )$, $D_{\rm visc} = 5.6 \sigma^2 / \tau$,
and $\tau_{\rm visc} = 0.02 \tau$.  The structural memory kernel has
$C_{\rm struc} = 0.42 k_B T \tau^2 / \sigma^2$, $D_{\rm struc} = 0.25
\sigma^2 / \tau$, and $\tau_{\rm struc} = 0.8 \tau$. Here the natural
      {\ms (Brownian)} time scale is $\tau=\gamma \sigma^2/(k_BT)$,
      where $\sigma$ is the hard sphere diamater.
{\ms For standard colloids with diameter $\sigma=1 \,\mu\rm m$
  dispersed in water at $T=20^o\,\rm C$, using the Stokes-Einstein
  form for $\gamma= 3\pi\eta \sigma$ \cite{hansen2013} gives
  $\gamma=9.44 \cdot 10^{-13}\,\rm kg \, s^{-1}$, which yields
  $\tau=2.33\,\rm s$. The corresponding memory times are $\tau_{\rm
    visc}= 0.046 \,\rm s$ and $\tau_{\rm struc}=1.87 \,\rm s$,
  i.e.\ values that are well inside of an experimentally accessible
  range. Using larger colloids \cite{stopper2018exp,royall2007} scales
  up the values for the memory times accordingly; the particles used
  e.g.\ for the (two-dimensional) system of Ref.~\cite{stopper2018exp}
  are of size $\sigma=4.04\,\mu\rm m$, which correspondingly upscales
  the values for both memory times by a factor of four. }

In order to gain further insight into the nature of the relevant
forces, we follow Ref.~\cite{geigenfeind2020superdemixing} and rewrite
the internal forces that act in a binary mixture as consisting of a
non-selective (``differential'') force field ($\fv_{\rm int}$) and a
selective (``total'') force density ($\Gv_{\rm int}$), such that the
self and distinct internal forces density distributions can
respectively be expressed as
\begin{equation}
  \Fv_{\rm int}^{\rm self} = 
  \rho_{\rm self} \fv_{\rm int} + \Gv_{\rm int},
  \quad
  \Fv_{\rm int}^{\rm dist} =
  \rho_{\rm dist} \fv_{\rm int} - \Gv_{\rm int}.
    \label{EQFintBothFromfG}
\end{equation}
Using the new fields $\fv_{\rm int}$ and $\Gv_{\rm int}$ in the
partial force density balance \eqref{EQcurrentForceBalancePartial}
leads to equations of motion for the total and for the
``differential'' motion,
\begin{eqnarray}
  \gamma \vel &=& -k_BT\nabla \ln \rho + \fv_{\rm int},
  \label{EQofMotionNonSelective}\\
  \gamma \Jv_\Delta &=& -k_BT \nabla \rho_\Delta 
  +\rho_\Delta\fv_{\rm int}
  +2\Gv_{\rm int},
  \label{EQofMotionSelective}
\end{eqnarray}
where the differential van Hove current is $\Jv_\Delta=\Jv_{\rm
  self}-\Jv_{\rm dist}$, and the differential van Hove function is
$\rho_\Delta=\rho_{\rm self}-\rho_{\rm dist}$.  Solving the linear set
of equations \eqref{EQFintBothFromfG}
yields
\begin{equation}
  \fv_{\rm int} = \Fv_{\rm int}/\rho,
\quad
  \Gv_{\rm int} = 
  (\rho_{\rm dist} \Fv_{\rm int}^{\rm self}
  -\rho_{\rm self} \Fv_{\rm int}^{\rm dist})/\rho,
\label{EQfAndGintDefinition}
\end{equation}
which allows to obtain results for $\fv_{\rm int}(\rv,t)$ and
$\Gv_{\rm int}(\rv,t)$ [from the correlators on the right hand sides
  of \eqref{EQfAndGintDefinition}.
Due to the linearity of the transformations 
\eqref{EQfAndGintDefinition}, splitting into adiabatic and
superadiabatic contributions applies according to $\fv_{\rm
  int}=\fv_{\rm ad}+\fv_{\rm sup}$ and $\Gv_{\rm int}=\Gv_{\rm
  ad}+\Gv_{\rm sup}$.

Figure \ref{figResults}(c) presents results for the van Hove function
of hard spheres. Shown are the self and the distinct part, $\rho_{\rm
  self}(r,t)$ and $\rho_{\rm dist}(r,t)$, at three different
representative times $t/\tau=0.1, 0.3, 0.6$. The results are obtained
using event-driven Brownian dynamics (BD) computer simulations
\cite{scala2007}. We use $N=1090$ particles in a three-dimensional
simulation box of size $15\times 10\times 10 \sigma^3$. The sampling
is based on $10^6$ time steps of size $10^{-3} \tau$, and hence an
overall simulation time of $10^3 \tau$. Appropriate filling of
histograms of particle pair distances yields results for the van Hove
function.

At the early time, $t=0.1\tau$ (first column of
Fig.~\ref{figResults}), the van Hove function has moderately decayed,
as compared to its initial condition.
Over the course of time, cf.\ the results for $0.3\tau$ (middle
column) and $0.6\tau$ (right column), the self part broadens and its
height correspondingly decreases. The initial correlation hole in the
distinct van Hove function is gradually being filled. Besides these
transport processes, the initially pronounced oscillations at
distances $r\gtrsim\sigma$ decay.

We demonstrate the agreement of adiabatic and dynamical density
profiles in Fig.~\ref{figResults}(c), by showing the MC simulation
results obtained from equilibrium sampling of the adiabatic state,
i.e.\ of the system in which the external potential $V_{\rm ad}^{\rm
  self}(\rv)$ acts on the (single) self particle and $V_{\rm ad}^{\rm
  dist}(\rv)$ acts on the remaining $N-1$ particles. (As $N$ is large
enough, we do not expect that finite size effects are relevant.)
Apart from very small numerical artifacts, clearly the agreement of
dynamical and adiabatic density profiles is excellent.  Hence we trust
results for the adiabatic force densities (presented below), obtained
via~\eqref{EQFadFromEquilibriumForceBalance}.

Besides the self and distinct parts, we also show results for the
total van Hove function, $\rho=\rho_{\rm self}+\rho_{\rm dist}$, and
the differential van Hove function $\rho_\Delta=\rho_{\rm
  self}-\rho_{\rm dist}$ in Fig.~\ref{figResults}(c). Clearly the
spatial structuring of $\rho$ is much reduced upon disregarding the
self-distinct labelling. Nevertheless, as all particles in the system
are ultimately identical, and the self-distinct labelling was
introduced for mere book-keeping purposes, one might wonder whether
the physically most relevant phenomena are revealed or are rather
hidden by the labelling.

In Fig.~\ref{figResults}(d) we show results for the different
contributions to the self force density. As a reference, we plot the
ideal contribution, $-k_BT\nabla\rho_{\rm self}$, which tends to
spread the self peak in time. Here positive (negative) values both of
force fields and of force densities indicate the outward (inward)
direction. The adiabatic force density counteracts the ideal part, and
hence tends to stabilize the self density peak. The DFT results yield
very satisfactory results, as compared to the MC data, at all times
considered. The superadiabatic self force density supports the effect
of $\Fv_{\rm ad}^{\rm self}$, but it has longer range and larger
magnitude at later times. Except for a slight overestimation at
$0.1\tau$, the PFT reproduces this effect very well, and hence
provides a mechanism for the slowing down of the dynamics.

The contributions to the distinct force density, shown in
Fig.~\ref{figResults}(e), show more complex, oscillatory behaviour, at
both earlier times. The ideal force density is again directly related,
via the spatial derivative, to the distinct density, shown in
Fig.~\ref{figResults}(c).  The oscillations of the distinct density
profiles are hence imprinted into ideal force density and their effect
is to homogenize the density. As is the case for the self part, the
adiabatic force density counteracts this effect, and hence tends to
stabilize the density oscillations. The DFT results for $\Fv_{\rm
  ad}^{\rm self}$ are very satisfactory, with some underestimation
inside of the core, $r \lesssim \sigma$. The superadiabatic distinct
force density has complex spatial features. It tends to slow down the
decay of the spatial structure. At early times, the magnitude is
smaller than that of $\Fv_{\rm ad}^{\rm dist}$, but this relationship
changes at later times, where $\Fv_{\rm sup}^{\rm dist}$ becomes
dominant. Again, up to some deviations inside of the core, the PFT
describes $\Fv_{\rm sup}^{\rm dist}$ in very good agreement with the
BD data.

In Fig.~\ref{figResults}(f) we show results for the contributions to
the differential force density, $\Gv_{\rm int}$, as defined in
\eqref{EQfAndGintDefinition} and being relevant for the differential
equation of motion \eqref{EQofMotionSelective}. Within the PFT, we can
clearly identify that $\Gv_{\rm sup}$ is solely due to the drag
effect, i.e.\ the friction generated by the interflow of the self and
distinct components.
{\ms This result is relevant for the hard sphere dynamics at long
  times. Taking only the drag force as the dominant internal effect
  and balancing it with ideal diffusion, the long time self diffusion
  coefficient follows as $D_L=k_BT/(\gamma+ \rho_bC_{\rm
    drag})$. Within this approximation we obtain $D_L=0.38
  \sigma^2/\tau\,(=0.16\,\mu {\rm m}^2/s$ for $\sigma=1\,\mu\rm m$ and
  $\tau=2.33\rm s$ as above), in very reasonable agreement with our
  bare simulation result of $D_L=0.32\sigma^2/\tau\,(=0.14\,\mu\rm
  m^2/s)$.}  In contrast, the motion of the total van Hove function,
for which we show the relevant force fields in
Fig.~\ref{figResults}(g), is due to both compressional viscosity and
structural forces, with both complex spatial and temporal behaviour,
which are well captured by the PFT.  Crucially, while the details of
the superadiabatic force fields vary depending on the type of
dynamical situation considered, regarding these as arising from a
kinematic functional \eqref{EQPtexc} reveals their universal
characteristics.

{\ms

In conclusion, we have traced the mechanisms that govern the time
evolution of the van Hove function for hard spheres by identifying
three different and universal types of nonequilibrium force
contributions, all of which have been shown previously to be relevant
across a broad spectrum of nonequilibrium and driven systems. The
forces are due to i) drag of the tagged (``self'') particle against
the surrounding fluid of distinct particles, ii) volume (or ``bulk'')
viscosity due to the correlation shells undergoing
compressional-expansional flow, and iii) structural nonequilibrium
effects, which stabilize the spatial liquid structure against
decay. The power functional approximation generates all three types of
nonequilibrium force fields in quantitative agreement with Brownian
dynamics computer simulation results.  Our results hence demonstrate
intimate interrelationships between equilibrium and nonequilibrium
hard sphere properties.  }

It would be interesting to investigate in future work the relationship
of our findings to Rosenfeld's excess entropy scaling
\cite{rosenfeld1977,dyre2018excessEntropyScaling}, as much advanced by
Truskett and his coworkers \cite{mittal2006,pond2011}, to the
nonequilibrium Ornstein-Zernike framework
\cite{brader2013noz,brader2014noz}, as well as to the findings by Dyre
and coworkers on universality across systems with different
interparticle interaction potentials
\cite{dyre2016topicalReview,costigliola2019}.

\begin{acknowledgments}
We thank Daniel de las Heras and Sophie Hermann for useful
comments. This work is supported by the German Research Foundation
(DFG) via project number 317849184.
\end{acknowledgments}

\end{document}